# Edge and surface plasmons in graphene nanoribbons

Z. Fei[1,2], M. D. Goldflam[1], J.-S. Wu[1], S. Dai[1], M. Wagner[1], A. S. McLeod[1], M. K. Liu[3], K. W. Post[1], S. Zhu[4], G.C.A.M. Janssen[4], M. M. Fogler[1], D. N. Basov[1]

[1]Department of Physics, University of California, San Diego, La Jolla, California 92093, USA
[2]Department of Physics and Astronomy, Iowa State University, Ames, Iowa 50011, USA
[3]Department of Physics, Stony Brook University, Stony Brook, New York, 11790, USA
[4]Department of Precision and Microsystems Engineering, Delft University of Technology, Mekelweg 2, 2628 CD Delft, Netherland

**Abstract**
We report on nano-infrared (IR) imaging studies of confined plasmon modes inside patterned graphene nanoribbons (GNRs) fabricated with high-quality chemical-vapor-deposited (CVD) graphene on $Al_2O_3$ substrates. The confined geometry of these ribbons leads to distinct mode patterns and strong field enhancement, both of which evolve systematically with the ribbon width. In addition, spectroscopic nano-imaging in mid-infrared 850-1450 $cm^{-1}$ allowed us to evaluate the effect of the substrate phonons on the plasmon damping. Furthermore, we observed edge plasmons: peculiar one-dimensional modes propagating strictly along the edges of our patterned graphene nanostructures.

**Main text**

Surface plasmon polaritons, collective oscillation of charges on the surface of metals or semiconductors, have been harnessed to confine and manipulate electromagnetic energy at the nanometer length scale.[1] In particular, surface plasmons in graphene are collective oscillations of Dirac quasiparticles that reveal high confinement, electrostatic tunability and long lifetimes.[2-16] Plasmons in graphene are promising for optoelectronic and nanophotonic applications in a wide frequency range from the terahertz to the infrared (IR) regime.[16,17] One common approach to investigate plasmons is based on nano-structuring of plasmonic media.[12,18] Large area structures comprised of graphene nanoribbons (GNRs) and graphene nano-disks have been extensively investigated by means of various spectroscopies.[12-17] These types of structures are of interest in light of practical applications including: surface enhanced IR vibrational spectroscopy[19,20], modulators[21], photodetectors[22] and tunable metamaterials[23,24]. Whereas the collective, area-averaged responses of graphene nanotructures are well characterized, the real-space characteristics of confined plasmon modes within these nanostructures remain completely unexplored.

In this work, we performed nano-IR imaging on patterned GNRs utilizing an antenna-based nanoscope that is connected to both continuous-wave and broadband lasers[25] (Supporting Information). As shown in Figure 1a, the metalized tip of an atomic force microscope (AFM) is illuminated by IR light thus generating strong near fields underneath the tip apex. These fields have a wide range of in-plane momenta $q$ thus facilitating energy transfer and momentum bridging from photons to plasmons.[3-12] Our GNR samples were fabricated by lithography patterning of high quality CVD-grown graphene single crystals[26] on aluminum oxide ($Al_2O_3$) substrates (Supporting Information). As discussed in detail below, the optical phonon of $Al_2O_3$ is below $\omega = 1000$ cm$^{-1}$ (Figure S2), allowing for a wide mid-IR frequency region free from phonons.

In Figure 1b, we show the AFM phase image displaying arrays of GNRs with various widths (darker parts correspond to graphene). The representative near-field images of the GNRs with widths of $W = 480$, 380, 270 and 155 nm are shown in Figure 1c–f, where we plot the near-field amplitude $s(\omega)$ normalized to that of the bare $Al_2O_3$ substrate (Supporting Information). As documented in previous studies[4-12], $s(\omega)$ is a direct measure of the $z$-component electric field amplitude $|E_z|$ underneath the AFM tip. In our experiments, we kept $p$-polarized light beam incident along the ribbons (Figure 1a) to avoid direct excitation of resonance modes of ribbons.[12] The excitation frequency $\omega$ is set to be at 1184 cm$^{-1}$, far away from the surface optical phonons of $Al_2O_3$ (Figure S2). The boundaries of these ribbons are marked with white dashed lines. In Figure 1g-j we also plot line profiles taken perpendicular to GNRs along red dashed lines in Figure 1c-f. The vertical dashed lines in Figure 1g-j mark the boundaries of GNRs. The fringes in Figure 1c-f correspond to the peaks in line profile plotted in Figure 1g-j.

We begin with the widest ribbons ($W = 480$ nm, Figure 1c,g) where two bright fringes form close to the boundaries of the ribbons. In addition to these brightest fringes, we also observe fringes in the interior of the ribbons (more clearly seen in Figure 1g). The former ones (marked with red arrows in Figure 1g-j), which we referred to as principal fringes, show much higher amplitude compared to the inner oscillations (marked with green arrows in Figure 1g,h). In nano-IR experiments, fringes originate from constructive interference between tip-launched and boundary-reflected plasmons (Figure 1a).[3-12] Apart from these familiar patterns, we also observed plasmonic characteristics due to ribbon confinement as the ribbon width ($W$) shrinks. First, the two principal fringes move closer to each other with decreasing $W$ and consequently the total number of distinct fringes observable within GNRs decreases. Second, we witnessed a strong enhancement of the plasmon intensity when the two principal fringes merge together and become indistinguishable ($W = 155$ nm, Figure 1f,j). The single fringe thus-obtained is highly confined laterally with a full width at half maximum (defined by a pair of blue arrows in Figure 1j) of ~50 nm. The peak amplitude appears to be the sum of the two principal fringes, so the plasmon intensity $I(\omega) \sim |s(\omega)|^2$ is about 4 times higher compared to the fringes in wider ribbons or larger area graphene. The general aspects of the ribbon confinement effects described above are consistent with previous studies of tapered GNRs.[4,5,9] Nevertheless, the data of straight GNRs are extremely valuable for quantitative analysis of the entire tip-ribbon system (Supporting Information).

In order to explore the evolution of the fringe patterns with the frequency of IR radiation, we performed nano-imaging of the GNRs with both broadband and CW sources following an approach introduced in our recent study[27]. The broadband mapping is given in Figure 2a, where we plot a real-space line scan across the GNR in Figure 1c (along the red dashed line). For every given real-space position ($x$) along the line scan, we collected a spectrum covering a broad frequency range (850 - 1450 cm$^{-1}$) with a spectral resolution of 3 cm$^{-1}$. Therefore Figure 2a contains both real-space and spectroscopic information about the plasmonic modes in GNRs. As an example, we show in Figure 2b horizontal line cuts (black curves) taken directly from the broadband map (Figure 2a) at several discrete frequencies (marked with white arrows in Figure 2a). As a comparison, we plot in Figure 2c the line profiles taken with the CW laser source across the same ribbon at these frequencies. The broadband and CW profiles share similar gross characteristics despite the fact that the peaks in the broadband profiles are noticeably weaker compared to those in the CW profiles. We attribute the intensity weakening in the broadband data to the electron heating (~1500 K) by the pulses of the broadband source: an effect that enhances plasmon damping[28] as verified by our modeling (see discussion below). We emphasize that the broadband map shown in Figure 2a was taken with only two hyper-spectral scans with each scan covering the spectral width of about 400 cm$^{-1}$ and collected within 30 min. The wide spectral range and short acquisition time endow this technique with major advantages over discrete-frequency scans carried out with monochromatic lasers. To the best of our knowledge, the broadband line scans for graphene plasmons are demonstrated here for the first time.

The main features of Figure 2 above 1000 cm$^{-1}$ are the two principal fringes close to the ribbon boundaries (marked with vertical dashed lines). As the frequency varies from 1450 cm$^{-1}$ to 1000 cm$^{-1}$, the two principal fringes move inward and eventually merge together at the ribbon center. Moreover, the fringe intensity increases significantly with decreasing frequency until reaching 1000 cm$^{-1}$. At even lower frequencies, fringes become strongly damped and their intensity rapidly declines. Note that we vary the scales of vertical axes in Figure 2b in order to visualize all illuminating details of the plasmonic features. The gross features of the data are captured by our numerical analysis (blue curves in Figure 2b,c), where we calculate the near-field amplitude by modeling the AFM tip as a conducting spheroid (Supporting Information).[4,6,29] The key modeling parameter is the complex plasmon wavevector ($q_p$) of graphene that contains information about both the plasmon wavelength $\lambda_p = 2\pi/\text{Re}(q_p)$ and plasmon damping rate $\gamma_p = \text{Im}(q_p)/\text{Re}(q_p)$.

As introduced in our previous works[4,6], the above modeling method is quantitatively accurate and it allows us to extract the complex $q_p$ by directly fitting the plasmon fringe profiles. The fitting was performed on both the broadband (Figure 2b) and CW (Figure 2c) data sets. The outcome of the fitting are data points of $\text{Re}(q_p)$ (Figure 3a) and $\gamma_p = \text{Im}(q_p)/\text{Re}(q_p)$ (Figure 3b) at various excitation frequencies. In order to get the best fit, we used the same set of $\text{Re}(q_p)$ but higher $\gamma_p$ (Figure 3b) when modeling the broadband profiles. As discussed above, the higher damping in the broadband fitting is due to the electron heating by femtosecond laser pulses (Supporting Information).[28] Thus-extracted data points for $\text{Re}(q_p)$ are overlaid on top of the frequency ($\omega$) – momentum ($q$) dispersion diagram (Figure 3a). In this diagram we plot the imaginary part of the $p$-polarized reflection coefficient $\text{Im}(r_p)$ of the entire graphene/substrate system produced using the standard transfer matrix method[3]. The optical conductivity of graphene used here for the dispersion calculation is obtained with the random phase approximation method where the graphene Fermi energy ($E_F$) is the only parameter. In the dispersion color plot, the bright curves represent the plasmon mode and the best match with the experimental data point of $\text{Re}(q_p)$ is obtained when $E_F \approx 2700$ cm$^{-1}$.

The spectroscopic nano-imaging data (Figure 2) and the dispersion plot (Figure 3a) obtained for graphene on the $Al_2O_3$ substrate convey two main messages. First, graphene plasmons are well defined and strongly dispersive above 1000 cm$^{-1}$, where plasmon fringes were clearly observed. Second, graphene plasmons are strongly damped below 1000 cm$^{-1}$: a likely outcome of the proximity to the substrate phonon around 830 cm$^{-1}$ (Figure S2). The phonon response was directly investigated using nano-IR spectroscopy (Figure 3c) that reveals a strong resonance around 830 cm$^{-1}$ (black curve in Figure 3d) consistent with the optical constants of $Al_2O_3$ obtained from ellipsometry (Figure S2). From Figure 3c, one can also see that graphene significantly enhances the phonon resonance and causes a slight blue shift (~7 cm$^{-1}$), which are both signature effects of the plasmon-phonon coupling.[3]

The direct outcome of the plasmon-phonon coupling is the strong damping of graphene plasmons[30] that leads to strongly damped GNR fringes observed below 1000 cm$^{-1}$ in Figure 2. For the purpose of quantitative analysis, we plot in Figure 3b the extracted plasmon damping rate $\gamma_p = \text{Im}(q_p)/\text{Re}(q_p)$ from fitting for both the CW and broadband excitations. In the case of CW excitation (black squares), one can see that $\gamma_p$ exceeds 0.2 within the plasmon-phonon coupling region (below 1000 cm$^{-1}$) and can reach values above 0.4 close to the $Al_2O_3$ phonon at 830 cm$^{-1}$ (Figure 3c). Away from this region (above 1100 cm$^{-1}$), $\gamma_p$ drops rapidly to 0.12, which is comparable to and even slightly lower than what was reported in previous studies conducted for exfoliated graphene on $SiO_2$ substrates ($\gamma_p \sim$ 0.135).[4] The plasmon damping rate could be further suppressed by thermal annealing and adding top protection layer to approach the recent experimental record ($\gamma_p \sim 0.025$) set by graphene encapsulated by hexagonal boron nitride.[9,31] When graphene and the AFM tip are illuminated by the broadband source (red circles), $\gamma_p$ is enhanced by a factor of two or three due to the electron heating by the femtosecond pulses. Note that the damping analysis performed here in our imaging study is directly related to the propagation length of graphene plasmons, while those done in previous spectroscopy work[14] are linked to the mode lifetime. As detailed in the Supporting Information, the two methods provide complementary views about graphene plasmon damping close to the substrate phonons.

Now we wish to discuss plasmonic responses at the edges of our structures. In Figure 4a, we show near-field images of a cross-cut GNR ($W = 480$ nm) that reveals two sharp corners at the top. The dominant features of these images are again the bright plasmon fringes parallel to the boundaries as discussed above. Hereafter we refer to these conventional plasmons as 2D surface plasmons. Remarkably, we also see in these images faint oscillations distributed *along* the graphene edges. Compared to the fringes of 2D plasmons, these latter oscillations confined to the edges are much weaker in intensity, and therefore can be easily overlooked. In order to reveal these oscillations along the edges, we adjust the color scales to maximize the contrast in all panels of Figure 4a.

We assert that the faint oscillations in Figure 4a are interference patterns due to *edge plasmons* of graphene. Such 1D excitations are generic for 2D metals although their dispersion depends on the charge density profile and the dielectric environment of the edge. In the simplest case where this profile is step-like and the conducting sheet resides at the interface of two uniform media, the wavelengths of the 1D edge plasmons ($\lambda_{ep}$) and 2D surface plasmons ($\lambda_p$) differ by a universal numerical factor[32,33]

$$\lambda_{ep} = 0.906 \lambda_p. \tag{1}$$

The physical reason for the smaller wavelength of the edge plasmons compared to the 2D ones (Eq. 1) is the effective reduction of the Drude weight at the edges, where free carriers exist only on "one side." The electric field of the edge plasmons decays exponentially away from the edge.

Similar to 2D surface plasmons, the 1D edge modes are launched by the AFM tip and can be bounced back by reflectors or scatterers. In Figure 4a, the sharp corners of the GNR serve as effective reflectors for the edge modes. For the purpose of quantitative analysis, we plot in Figure 4b line profiles taken along the red and black dashed lines in Figure 4a. Thus, the red profile reveals the edge oscillations (red curves) and whereas the black traces visualize fringes of 2D surface plasmons. We first focus to the top panel of Figure 4b, where the excitation frequency is set to be $\omega = 1160$ cm$^{-1}$. A quick inspection shows that the profile of the edge oscillations shares a similar line shape with those of the fringes of 2D plasmons despite their huge intensity difference. Both types of profiles reveal a number of peaks with two strongest ones (principal peaks, marked with arrows) close to the left and right reflectors (marked with dashed lines). As discussed in detail above, such a line shape is a signature of mode confinement between the left and right reflectors. Despite the above similarities, a closer inquiry into these profiles uncovers a number of differences. Specifically, the positions of the two principal peaks of the edge oscillations are slightly closer to the reflectors (black dashed lines). Also, more peaks appear in the profiles of the edge oscillations compared to those of the 2D plasmonic fringes. Indeed, we observed five peaks for the edge modes versus three peaks for the surface plasmon modes at $\omega = 1160$ cm$^{-1}$. The above mentioned observations are consistent with the notion of smaller wavelength of the edge modes compared to 2D surface plasmons. A quantitative estimation can be obtained by measuring the distances from the left and right principal peaks (marked with arrows) to their adjacent reflectors (dashed lines): $D_L$ and $D_R$. The average value of the two measured distances $(D_L+D_R)/2$ is roughly proportional to the plasmon wavelength. Take $\omega = 1160$ cm$^{-1}$ for example, $(D_L+D_R)/2$ is 77 nm and 95 nm for edge and surface modes respectively, so the ratio between plasmon wavelengths of the two modes $\lambda_{ep}/\lambda_p$ at this frequency is about 0.81. Similarly, we can estimate the ratios $\lambda_{ep}/\lambda_p$ to be 0.81, 0.82 for $\omega = 1100$ and 1065 cm$^{-1}$ respectively. These values roughly agree with the expectations of Eq. 1. The small deviation (~10%) might be due to a slightly different doping at the edges compared to the interior of graphene. Our hypothesis of edge plasmons is further verified by the frequency-dependence study of these edge oscillations (Figure 4b). From $\omega = 1160$ cm$^{-1}$ to 1065 cm$^{-1}$ (Figure 4b), we found that the two principal peaks of edge mode profiles (red) move further away from the two corners and the total number of edge oscillations decreases. Both observations indicate an increased $\lambda_{ep}$ with decreasing frequency, which is consistent with the dispersion nature of edge plasmons.

In fact, oscillations due to edge plasmons are not only observed close to sharp corners of GNRs. These oscillations appear in nearly all lithography-patterned graphene structures, but are rarely seen in exfoliated graphene flakes with natural, non-processed edges. Lithography patterning presumably creates much more edge defects (or roughness) compared to exfoliation. Although these defects are not clearly identified by our AFM possibly due to their small sizes, they could still serve as plasmon reflectors or scatterers (Figure S3a): a scenario analogous to atomic-scale grain boundaries in graphene[6]. As an example, in Figure 4c we plot data for a patterned graphene micro-disk; signal oscillations surrounding the disk are seen in this image. In Figure 4d, we show zoom-in views of the micro-disk taken at several IR frequencies from the region defined by the red rectangle in Figure 4c. Here, there are three bright modes (marked with white arrows) within the field of view. Unlike the edge modes confined by the two sharp corners discussed above (Figure 4a,b), the positions of the modes surrounding the micro-disk (or any other patterned structures away from sharp corners) do not show obvious frequency dependence. We believe these fixed-position modes are signatures of plasmon localization due to a high density ($\alpha$) of reflectors (edge defects here) per unit length. As demonstrated in Figure S4 (Supporting Information), strong localization could occur when $1/\alpha$ is much smaller than $\lambda_{ep}$ and consequently the distribution of the localized modes is solely determined by the positions of the reflectors. Future studies combining atomic-resolution characterization tools with nano-IR imaging are necessary to correlate the structural properties and plasmonic responses of these edge defects. Note that edge defects possibly also exist close to the sharp corners in Fig. 4a, but they are much weaker plasmon reflectors compared to the corners due to their small sizes. As a result, sharp corners play a dominant role in the formation of nearby edge oscillations. Therefore, we didn't consider the effects due to defects in our analysis of the data of sharp corners in Fig. 4a,b.

For completeness, we wish to mention several previous reports of edge plasmons in graphene nanostructures using optical and microwave spectroscopy.[34-36] These studies were facilitated by a drastic suppression of plasmonic losses in the presence of a quantizing magnetic field (note that Eq. 1 does not apply in such conditions[32,34-36]). In contrast, the propagation length of the zero-field edge plasmons is not expected to exceed much that of the 2D plasmons. Zero-field edge plasmons have been however detected in other systems, e.g., metallic nanoplatelets, using electron energy-loss spectroscopy.[37,38] To the best of our knowledge, Figure 4 is the first visualization of the edge plasmons carried out by direct nano-IR imaging.

In conclusion, we have systematically studied the confined plasmon modes inside GNRs on the $Al_2O_3$ substrates. Through IR nano-imaging experiments, we observed intriguing plasmon fringe patterns that evolve systematically with both the ribbon width and the excitation IR frequencies. In addition, confined geometry of GNRs leads to an enhancement of the plasmon field. Moreover, we demonstrate by spectroscopic imaging that substrate phonons play a major role in plasmon damping, and that $Al_2O_3$ offers a wide phonon-free region in the mid-IR where the plasmon damping is significantly reduced. Finally, we observed real-space signatures due to edge plasmons in graphene. Our numerical simulations suggest that distinct mode patterns due to both edge plasmons and bulk plasmons should appear in devices with smooth graphene edges (Figure S3c).

**Associated Content**
Supporting Information: Details of the experimental setup, sample preparation, data analysis and modeling.
This material is available free of charge via the Internet at http://pubs.acs.org .

**Author Information**
Corresponding Author
*Email: (Z.F) zfei@iastate.edu .
Notes
The authors declare no competing financial interest.

**Acknowledgment**
Authors acknowledge support from ONR and AFOSR. The development of scanning plasmon interferometry is supported by DOE-BES and ARO. DNB is supported by the

Gordon and Betty Moore Foundation's EPiQS Initiative through Grant GBMF4533. S.Z. and G.C.A.M.J acknowledge the financial support from the Young Wild Idea Grant of the Delft Centre for Materials and Delft Energy Initiative Fund, as well as Foundation for Fundamental Research on Matter (FOM) and Netherlands Organisation for Scientific Research (NWO).

**Figure Captions**

**Figure 1**. Nano-IR imaging of confined plasmons inside GNRs. (a) Schematic of a nano-IR imaging experiment. A metalized AFM tip is illuminated with IR light launching plasmons waves inside a GNR on the $Al_2O_3$ substrate. (b) The AFM phase image of GNR arrays with various widths. The darker parts of the AFM phase image correspond to graphene. (c-f) Near-field data images of GNRs with widths $W$ = 480, 380, 270 and 155 nm at $\omega$ = 1184 cm$^{-1}$. White dashed lines mark the boundaries of GNRs. All data were acquired at ambient conditions. (g-j) Line profiles taken perpendicular to GNRs along the red dashed lines in c-f. Black dashed lines mark the boundaries of the ribbons. The red/green arrows mark the positions of the principal/inner fringes. The blue arrows in (j) mark the full width at half maximum of the single fringe.

**Figure 2**. Spectroscopic nano-imaging of the confined plasmonic modes inside GNRs. (a) Broadband map of IR near-field amplitude $s(x, \omega)$ taken with a broadband pulse laser source along the red dashed line in Figure 1c. (b) Horizontal line cuts (black curves) taken from the broadband map at several discrete IR frequencies ($\omega$ = 890, 935, 1065, 1120 and 1184 cm$^{-1}$) marked with white arrows in panel (a). The blue curves display modeling results obtained as described in the text. (c) Line profiles (black curves) of IR near-field amplitude $s(\omega)$ taken with CW lasers at several discrete frequencies $\omega$ = 890, 935, 1065, 1120 and 1184 cm$^{-1}$ along the red dashed line in Figure 1c. The blue curves display modeling results obtained as described in the text. At all panels, veridical dashed lines mark the boundaries of the GNR and near-field amplitude $s(\omega)$ is normalized to that of the $Al_2O_3$ substrate.

**Figure 3. a**, The dispersion plot of graphene plasmons on the $Al_2O_3$ substrate, where we plot the imaginary part of the reflection coefficient $r_p(q, \omega)$ with a chemical potential of $|\mu|$ = 2700 cm$^{-1}$. The purple data points are extracted $Re(q_p)$ at various excitation frequencies obtained by fitting the fringe profiles in Figure 2b,c. **b**, The damping rate $\gamma_p$ of graphene plasmons extracted by fitting both the CW and broadband profiles in Figure 2b,c, respectively. **c**, Nano-IR spectra of bare $Al_2O_3$ substrate and large-scale graphene on $Al_2O_3$.

**Figure 4**. Nano-IR imaging of edge plasmons. **a**, Near-field images of a cross-cut GNR (lateral width $W$ = 480 nm) at three different IR frequencies. White dashed lines mark the boundaries of the cross-cut GNR. **b**, Line profiles of 2D surface plasmon modes (along the black dashed line) and 1D edge modes (along the red dashed line) taken directly from **a**. Dashed lines mark the left and right boundaries of the ribbon. The arrows mark the principal peaks in the profiles. **c**, Near-field image of a patterned graphene micro-disk taken at $\omega$ = 1160 cm$^{-1}$. **d**, Near-field images of a zoom-in region of **c** (defined by the red rectangle) taken at four different IR frequencies.

Figure 1

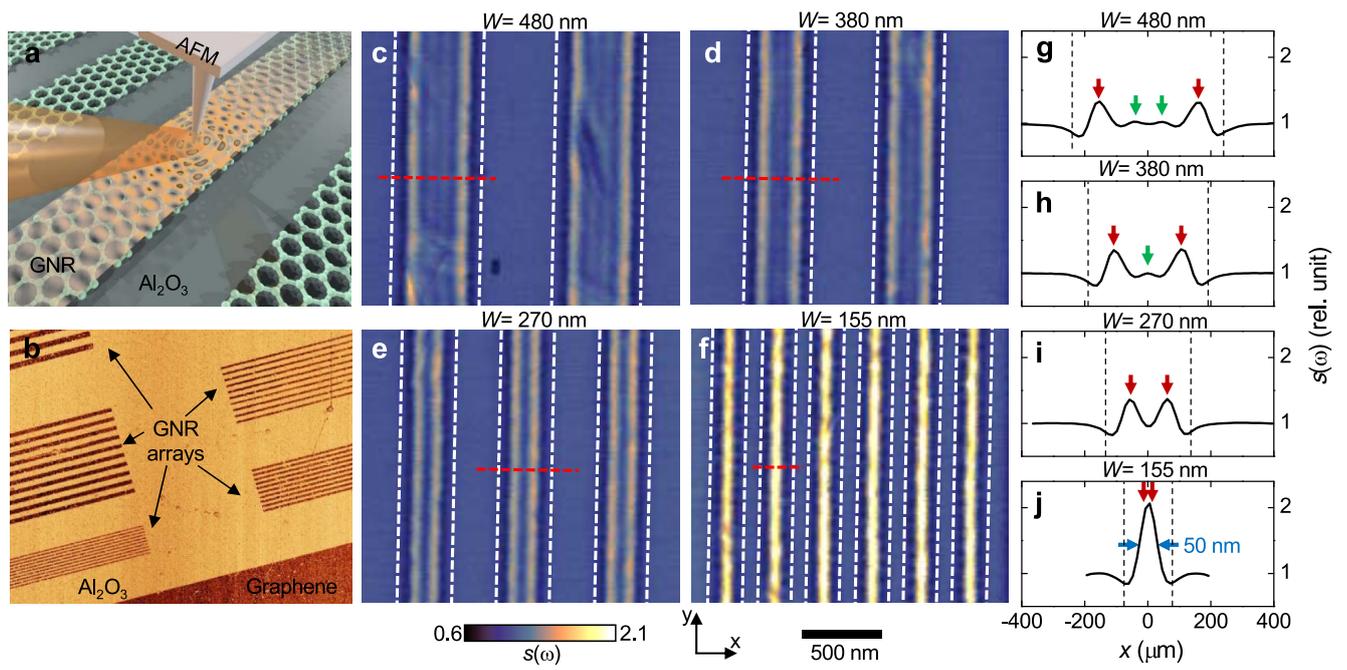

Figure 2

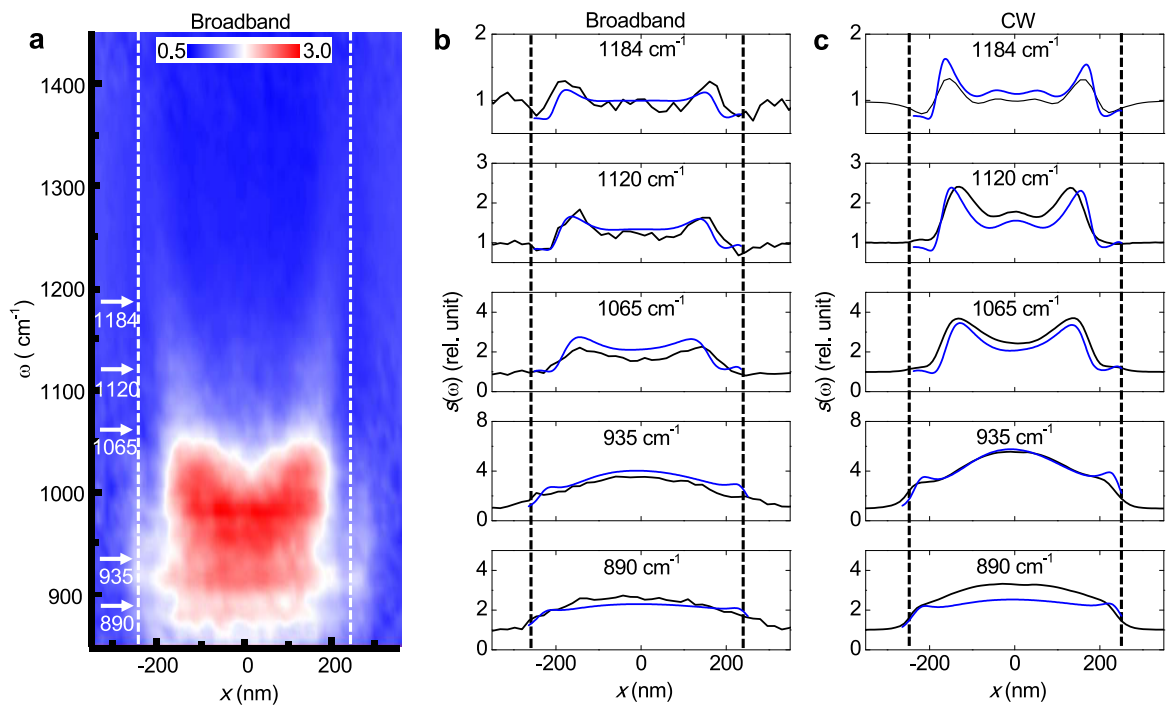

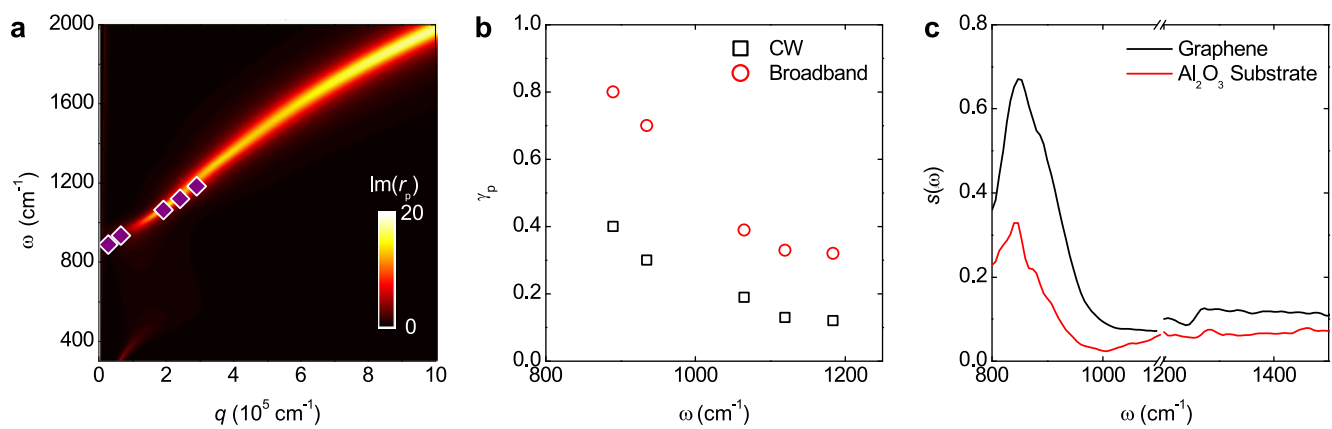

Figure 4

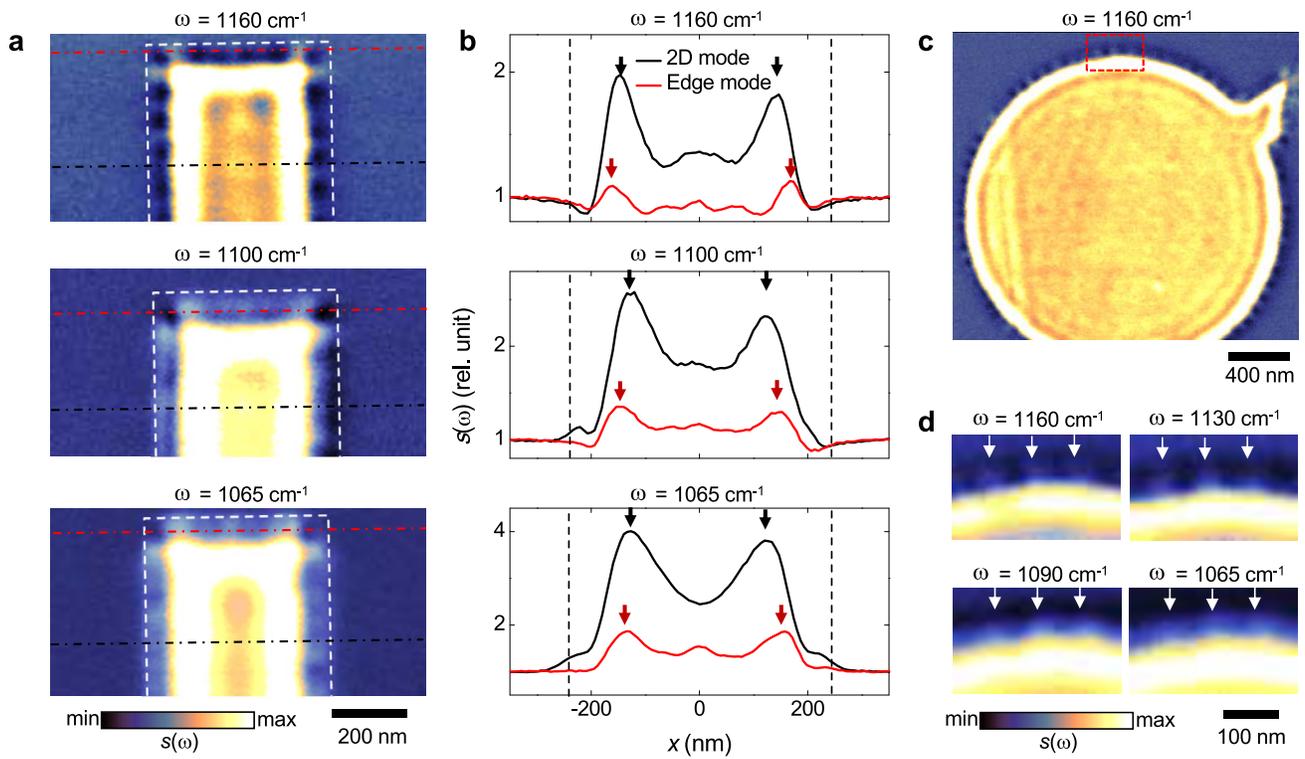

# Supporting Information of
# "Edge and Surface Plasmons in Graphene Nanoribbons"

Z. Fei*[1,2], M. D. Goldflam[1], J.-S. Wu[1], S. Dai[1], M. Wagner[1], A. S. McLeod[1], M. K. Liu[3], K. W. Post[1], S. Zhu[4], G.C.A.M. Janssen[4], M. M. Fogler[1], D. N. Basov[1]

[1]Department of Physics, University of California, San Diego, La Jolla, California 92093, USA
[2]Department of Physics and Astronomy, Iowa State University, Ames, Iowa 50011, USA
[3]Department of Physics, Stony Brook University, Stony Brook, New York, 11790, USA
[4]Department of Precision and Microsystems Engineering, Delft University of Technology, Mekelweg 2, 2628 CD Delft, Netherland

*Email: (Z.F.) zfei@iastate.edu

## 1. Experimental Setup

In order to perform nano-infrared (IR) imaging of graphene plasmons, we employed the scattering-type scanning near-field optical microscopy (s-SNOM). Our s-SNOM apparatus (Neaspec) is based on an atomic force microscope (AFM) operating in the tapping mode. Measurements were acquired at an AFM tapping frequency of $\Omega \approx 270$ kHz and a tapping amplitude of about 60 nm. As illustrated in Figure 1a, we utilized a metalized AFM probe, which is illuminated by a $p$-polarized mid-IR beam from either continuous wave (CW) lasers (Daylightsolutions) or broadband sources. The latter is based on the difference-frequency-generation (DFG) technique described in detail in Ref. 1. The broadband pulses generated by the DFG system has a pulse width of about 100 fs and they are continuously tunable in a wide mid-IR range (650 - 2400 cm$^{-1}$). The illuminated AFM tip (Figure 1a in the main text) generates strong near fields with a wide range of in-plane momenta $q$.[2] These momenta could exceed the far-field wavevector by two orders of magnitude, thus allowing for energy transfer and momentum matching between the incident photons and excited plasmons.[2] The standard observable of an s-SNOM experiment is $s$: the near-field scattering amplitude demodulated at the $n^{\text{th}}$ ($n = 2$ in the current work) harmonics of the AFM tip oscillation.

## 2. Sample preparation

Our graphene nanoribbons (GNRs) were patterned by electron beam lithography on high quality graphene single crystals grown by the chemical vapor deposition (CVD) method.[3] The mobility of these single crystals is comparable to exfoliated microcrystals[4] and their typical sizes exceed 100 μm (Figure S1), sufficient for a wide range of technological applications. These CVD graphene crystals were then transferred to aluminum oxide ($Al_2O_3$) substrates grown by atomic layer deposition on silicon wafers. In order to determine the optical properties of $Al_2O_3$, we performed mid-IR ellipsometry on our wafers. Thus obtained dielectric function of $Al_2O_3$ is given in Figure S2, where one can see an optical phonon resonance below 1000 cm$^{-1}$. The corresponding transverse (TO), surface (SO) and longitudinal (LO) phonon frequencies are at $\omega = 629$, 830 and 924 cm$^{-1}$, respectively. The near-field resonance of $Al_2O_3$ substrate shown in the nano-FTIR plot (Figure 3c in the main text) is at the surface optical phonon frequency.

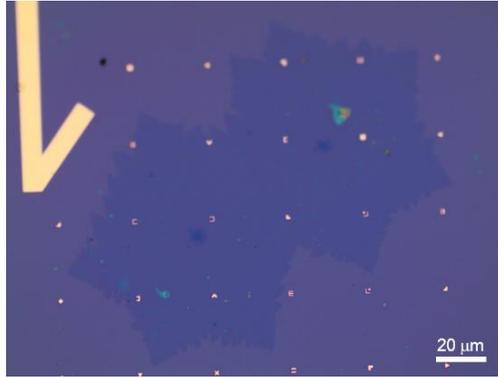

**Figure S1.** Optical microscopy image of a typical large area CVD graphene microcrystal. The substrate here is the standard $SiO_2$/Si wafer that gives the best optical contrast.

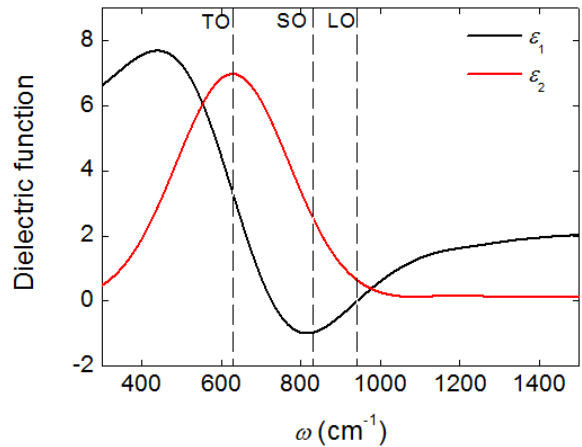

**Figure S2.** Real ($\varepsilon_1$) and imaginary ($\varepsilon_2$) parts of the dielectric function of $Al_2O_3$ determined through ellipsometry. The dashed lines mark respectively the transverse optical phonon (TO), surface optical phonon (SO) and longitudinal optical phonon (LO) frequencies.

### 3. Numerical modeling of plasmon fringe profiles

In order to model the fringes profiles of two-dimensional (2D) surface plasmons confined inside GNRs, we model our AFM tip as a metallic spheroid: the length of the spheroid is $2L$ and the radius of curvature at the tip ends is $a$. Here, $a$ is set to be 25 nm according to the manufacturer and $L$ is not a sensitive parameter so long as $L \gg a$. The scattering amplitude $s$ (before demodulation) scales with the total radiating dipole ($p_z$) of the spheroid. Therefore, in order to fit the line profiles perpendicular to the fringes inside GNRs, we need to calculate $p_z$ at different spatial coordinates ($x$, $z$) of the lower end of the AFM tip. Here, $x$ is the in-plane coordinate perpendicular to the GNRs and $z$ is the out-of-plane coordinate parallel to the AFM tip. By calculating $p_z$ at different $z$, we are able to perform 'demodulation' of the scattering amplitude $s$ and get different harmonics of the scattering signal. While calculating $p_z$ at different $x$ allows us to plot the modeling profiles of near-field amplitude. The symmetry along the ribbon direction ($y$ direction) of our straight GNRs saves us orders of magnitude time and efforts in modeling the plasmon fringes profiles, which makes it practical to fit the essential parameters of graphene plasmons by exploring the broad parameter space. In the current work, the scattering amplitude $s$ is normalized to that of bare $Al_2O_3$ substrate, so it has a relative unit (rel. unit). The dielectric function $\varepsilon_1(\omega) + i\varepsilon_2(\omega)$ of $Al_2O_3$ (Figure S2) we used for the modeling was determined by ellipsometry.

## 4. Damping of graphene plasmons

In the main text, we introduced our real-space analysis of graphene plasmon damping close to the substrate phonon. In fact, plasmon damping has also been discussed in previous spectroscopic study of GNRs.[5] Here, we want to discuss in detail the two methods and their conclusions. In the spectroscopic work[5], damping analysis is done based on linewidth analysis of plasmon resonance peaks of GNRs. The damping rate obtained in this way (labeled as $\gamma_1$) is thus inversely proportional to resonance lifetime. In our nano-IR imaging work, damping analysis was done by fitting the real-space data and thus the extracted damping rate (labeled as $\gamma_2$) is inversely related to the mode propagation length. The outcomes of the two analyses are not equivalent and are seemingly "contradictory" to each other. For example, according to Ref. 5, the damping rate ($\gamma_1$) is lower close to the substrate phonon. Nevertheless, data and analysis in the current work conclude that the damping rate ($\gamma_2$) is much higher close to the substrate phonon. In order to explain the "contradiction", we need to understand the relation the two damping rates: $\gamma_1$ to $\gamma_2$. They are of course not equal to each other as they are measured in different units. One has to consider the mode velocity ($v_p = d\omega/dq$) to connect the two damping rates: $\gamma_1 \sim v_p \gamma_2$. Close to the substrate phonon, the dispersion curve becomes flat ($v_p \rightarrow 0$), so $\gamma_1$ drops despite of an increased $\gamma_2$. Therefore, our real-space imaging study provides a complimentary view on graphene plasmon damping.

## 5. Further discussion of edge plasmons

Edge plasmons are one dimensional (1D) modes propagating strictly along the edges of 2D conductors such as graphene. Similar to the conventional 2D surface plasmons, the edge plasmons are launched by the near-field probe of the s-SNOM. These edge modes could be reflected back when encountering reflectors thus causing interference, which can be captured by the s-SNOM as edge oscillations. We discussed two types of reflectors in the main text of the paper. Type-1 reflectors are sharp corners (Fig. 4a,b) that can be seen clearly from AFM topography. Type-2 reflectors (Fig. 4c,d and also illustrated in Fig. S3a) are possibly small edge defects created by lithography patterning that are not easily identified by AFM. Type-1 reflectors have well defined positions. Moreover, they are much more efficient reflectors compared to the latter ones, thus generating higher-intensity edge oscillations that dominate over those due to type-2 reflectors. Therefore, we chose edge oscillations close to the sharp corners for quantitative analysis (Fig. 4a,b). Indeed, the edge oscillations between two separated sharp corners show obvious frequency dependence (Fig. 4b), consistent with the dispersion nature of edge plasmons.

The edge oscillations due to small edge defects (e.g. around a micro-disk, Fig. 4c,d and Fig. S3a) are much more difficult to analyze quantitatively. They may depend on the plasmon wavelength as well as the locations, sizes and shapes of these defects. We have carefully examined both the topography and near-field images of our data. Future studies combining high-resolution techniques (e.g. STM) with nano-IR imaging are necessary to uncover the properties of these defects.

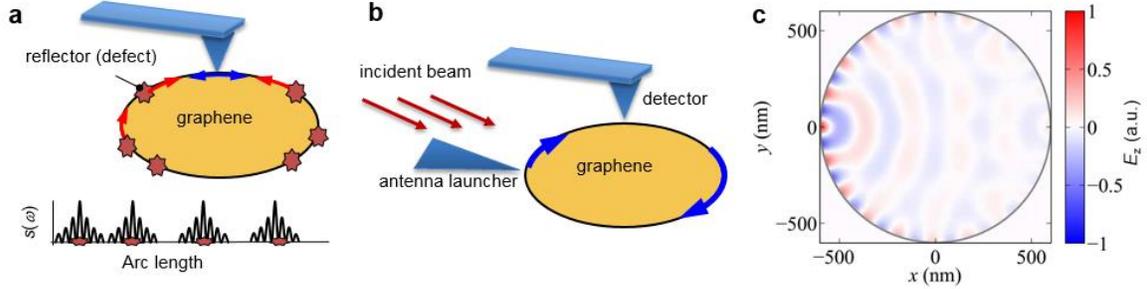

**Figure S3**. **a**, Qualitative explanation of the edge oscillations in Figure 4c,d of the main text as a superposition of many standing waves created by edge plasmons reflected off multiple random defects (stars). **b,** The proposed scheme to directly launch and detect edge plasmons. **c**, Numerically calculated $E_z$ field map excited inside a graphene disk by a metallic antenna. The antenna has a triangular shape (shown schematically in **b**) and its sharp tip touches the disk at point (0, 0). The excited 1D edge modes and 2D surface plasmons are visible.

## 6. Numerical modelling of edge plasmons

The physics of the zero-field edge plasmons in graphene may be explored in future studies using samples with clean, natural edges. The launching of edge modes can be achieved with the help of antenna-like structures (Figure S3b). As illustrated in Figure S3b, when the metallic sharp antenna is illuminated by the incident IR beam, the strong confined near fields close to the end of the antenna could launch plasmon waves inside adjacent graphene structure with smooth graphene edges. Our numerical simulations suggest that distinct mode patterns due to both 1D edge plasmons and 2D surface plasmons should appear in graphene (Figure S3c). As seen from Figure S3c, the wavelength of edge plasmons is smaller than that of surface plasmons in the graphene interior, which is consistent with eq. 1 in the main text. One can in principle map such mode pattern by using the s-SNOM.

We were unable to determine exactly the locations, sizes and shapes of the small edge defects due to the limited resolution of the AFM. We speculate that the edge modes are localized based on the fact that the oscillations due to these defects do not show obvious frequency dependence (Figure 4d in the main text). As detailed below, localization of edge plasmons could occur when the density ($\alpha$) of the reflectors (defects) per unit length is so high that their reflector-to-reflector separation (~ $1/\alpha$) is smaller than the wavelength of edge plasmons ($\lambda_{ep}$). A simple illustration is given in Figure S4, where we consider two defects with a separation of $d \sim 1/\alpha$ along the graphene edge (Figure S4a). When tip-launched edge plasmons encounter the reflectors, they are partially reflected and also undergo a $\sim \pi$ phase shift. In Figure S4b, we show the ($x_{tip}$, $\lambda_{ep}$) dependent plasmon amplitude ($|E_z|$) map, where $x_{tip}$ is the location of the AFM tip. Every horizontal line cut of the map resembles what we measure in our nano-IR imaging experiment at a given IR frequency. The two white dashed lines mark the positions of the reflectors. One can see that when $\lambda_{ep}/d \leq 1$, there are two or more bright modes confined within the two reflectors, which is consistent with our observations of edge modes confined between two sharp corners (Figure 4a,b). However, as $\lambda_{ep}/d$ increases above 1.5, there is only single bright mode appears at the center of the two reflectors. The position of this single bright mode does not change with $\lambda_{ep}$ (or IR frequency) consistent with our observation of the fix-position modes surrounding the micro-disk (Figure 4d). Moreover, we can also see that the intensity of the single plasmon mode between the two reflectors decreases with increasing $\lambda_{ep}$ (or decreasing IR frequency), which is also consistent with our experimental observation (Figure 4d).

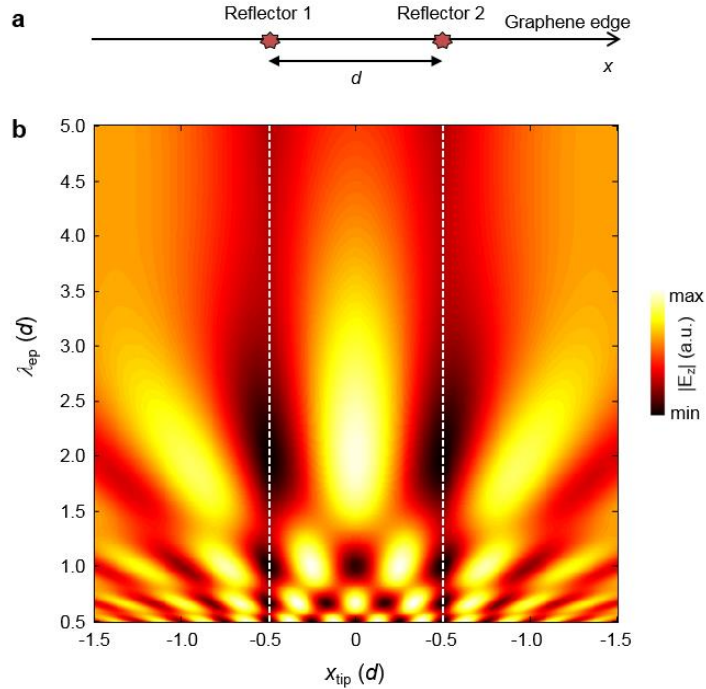

**Figure S4. a**, Illustration of a model for simulation of two reflectors (defects) at the graphene edge. **b**, Simulated ($x_{tip}$, $\lambda_{ep}$) dependent amplitude ($|E_z|$) map of edge plasmons created by the two reflectors in **a**.